\documentclass[prx,reprint,superscriptaddress,twocolumn,showkeys]{revtex4-1}
\usepackage[utf8]{inputenc} 
\usepackage{amsmath}
\usepackage{braket}
\usepackage{graphicx}
\usepackage{pgfplots}
\usepackage[T1]{fontenc}
\usepackage{upgreek}
\usepackage{csquotes}
\usepackage{comment}
\usepackage{natbib}
\usepackage{hhline}
\usepackage{amssymb}

\usepackage{dcolumn}
\usepackage{tabularx}
\usepackage{hyperref}
\usepackage{float}
\usepackage{listings}
\usepackage{color}
\definecolor{mygreen}{rgb}{0,0.6,0}
\definecolor{mygray}{rgb}{0.5,0.5,0.5}
\definecolor{mymauve}{rgb}{0.58,0,0.82}

\newcolumntype{C}{>{\centering\arraybackslash}X}
\pgfplotsset{compat=1.14}
\begin{document}
\title{Digital Quantum Simulation of Laser-Pulse Induced Tunneling Mechanism in Chemical Isomerization Reaction}

\author{Kuntal Halder}
\email{kuntal.h123@gmail.com}
\affiliation{Department of Physics,\\ Indian Institute of Technology Madras, Chennai, 600036, India}

\author{Narendra N. Hegade}
\email{narendrahegade5@gmail.com}
\affiliation{Department of Physics,\\ National Institute of Technology Silchar, Silchar, 788010, India}

\author{Bikash K. Behera}
\email{bkb13ms061@iiserkol.ac.in}
\author{Prasanta K. Panigrahi}
\email{pprasanta@iiserkol.ac.in}
\affiliation{Department of Physical Sciences,\\ Indian Institute of Science Education and Research Kolkata, Mohanpur, 741246, West Bengal, India}

\begin{abstract}
Using quantum computers to simulate polyatomic reaction dynamics has an exponential advantage in the amount of resources needed over classical computers. Here we demonstrate an exact simulation of the dynamics of the laser-driven isomerization reaction of asymmetric malondialdehydes. We discretize space and time, decompose the Hamiltonian operator according to the number of qubits and use Walsh-series approximation to implement the quantum circuit for diagonal operators. We observe that the reaction evolves by means of a tunneling mechanism through a potential barrier and the final state is in close agreement with theoretical predictions. All quantum circuits are implemented through IBM's QISKit platform in an ideal quantum simulator.
\end{abstract}


\maketitle

{\emph{\textbf{Introduction}}: 
Feynman first suggested in his seminal lecture in 1981 \cite{Feynman1982} that a complex quantum system can be efficiently simulated with another quantum system on which we have much greater control on. This new framework for simulating physics offers exponential advantage in the resources and time required over classical computers in most of the quantum systems. While at least thousands of qubits are required to solve classically difficult problems such as prime factorization and ordered searching, but few hundreds of qubits are enough to efficiently simulate quantum systems that are classically unfeasible \cite{Buluta2009}. A universal quantum simulator \cite{Lloyd1996} would be a device that operates on the principles of quantum mechanics and efficiently simulates the dynamics of any other many-body quantum systems with short range interactions. 

Classical simulation of chemical reaction dynamics is extremely challenging due to exponential growth in required resources with increasing degrees of freedom. In digital quantum simulation, the Hamiltonian of the system of interest is written as sum of local Hamiltonians,
the time evolution operator is decomposed by some approximation methods so that we can implement the dynamics of the system using quantum circuits.
In recent years, many algorithms have been proposed to simulate quantum chemistry problems
\cite{Aspuru-Guzik2005,Kassal2008,Lanyon2010,Kassal2011,Babbush2015,OMalley2016,Kandala2017,Kivlichan2018,Sim2018} for example finding molecular energy \cite{Colless2018}, chemical dynamics \cite{Kassal2008,Sornborger2018}, and their experimental realisation can be found here \cite{Lu2011,Peruzzo2014,Du2010,Hempel2018}.
In this article, we have shown the simulation of laser-induced tunneling phenomena in chemical isomerization reaction on a cloud-based quantum computing platform QISKit \cite{qiskit}. In the presence of laser field the hydrogen transfers from the reactant state to product state via tunneling through the potential barrier. To show this tunneling mechanism we have used three qubits to represent the potential energy curve on a 8 grid points \cite{Hegade2017}. The result we observed from the simulation is in good agreement with the theoretical prediction. The main motivation of this paper is to show the simulation of complex quantum dynamics on a today's quantum computer consisting of few number of qubits.} 

\begin{figure}[H]
   \begin{center}
   \includegraphics[width=\linewidth]{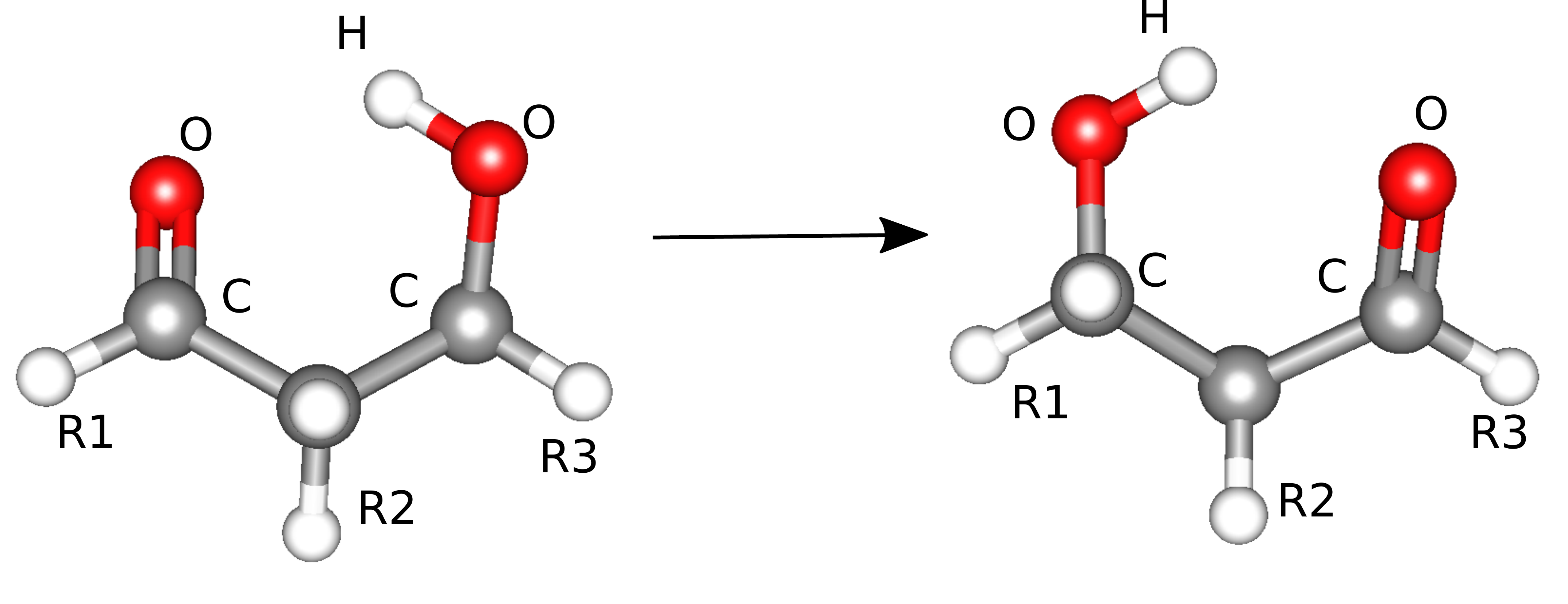}
    \caption{\textbf{Isomerization reaction of asymmetric malondialdehydes:} a hydrogen is transferred from one oxygen to another one.}
    \label{malondialdehyde}
    \end{center}
\end{figure}

{\em \textbf{Scheme:}}
There are different strategies that have been proposed in the literature to achieve laser-driven isomerization reaction.
Here we have considered the method called "hydrogen-subway" proposed by N. Doslic et al. \cite{Doslic1998} where the wave packet representing the reactant state in the presence of applied laser field tunnels through the potential barrier and gives the product. The reaction process is shown in Fig. \ref{malondialdehyde}.

For the simulation of laser-driven isomerization reaction in one dimension, we consider the time-dependent Schr$\ddot{\textrm{o}}$dinger's equation

\begin{equation}
i \hbar \frac{\partial \psi}{\partial t}= H_{mol} \psi+H_{int}\psi,
\end{equation}

where the molecular Hamiltonian $H_{mol}$ is given by 
\begin{equation}
    H_{mol}=T+V.
\end{equation}

Here T is the kinetic energy operator and the asymmetric double-well quatric potential $V$ can be implemented by the potential energy operator as a function of position
\begin{equation}
    V(x)=\frac{\Delta}{2x_0} (x-x_0)+ \frac{V_b-\Delta/2}{x_0^4} (x-x_0)^2 (x+x_0)^2
\end{equation}

\begin{figure}[H]
   \begin{center}
   \includegraphics[width=0.9\linewidth]{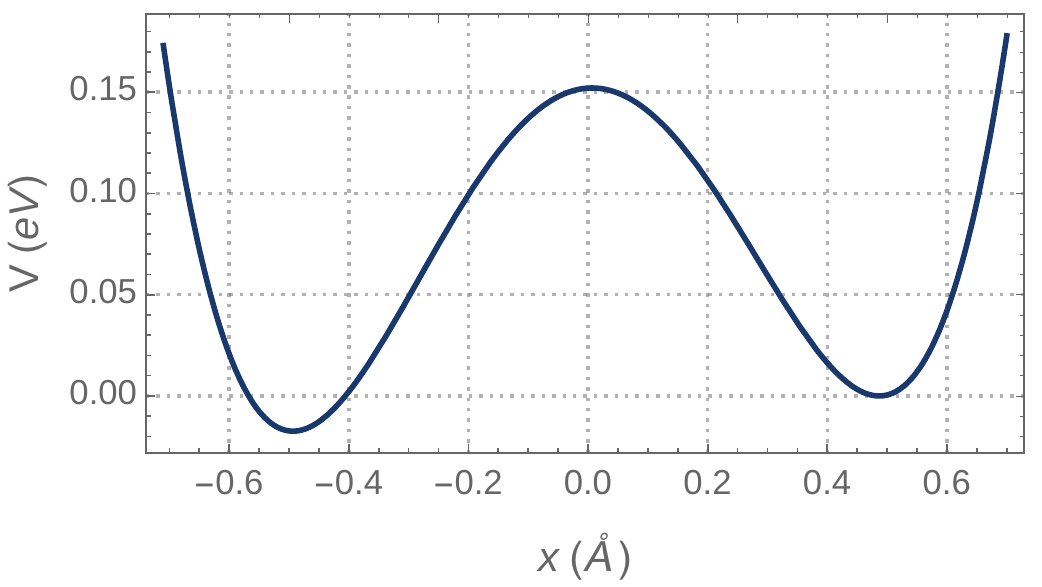}
   \end{center}
    \caption{\textbf{Asymmetric double-well quartic potential}. The reactant and product states are localized in separate wells.}
    \label{potential}
\end{figure}

where $\pm x_0$ gives the positions of the minima, $V_b$ represents the height of the barrier and the asymmetry of the wells can be changed by varying $\Delta$. 

The laser-molecule interaction Hamiltonian is
\begin{equation}
    H_{int}=E(t)=-\mu \epsilon(t)
\end{equation}
where $\mu=e x$ represents the dipole moment operator and $\epsilon(t)$ is the driving electric field.

The time evolution of the system is given by
\begin{equation}
    \ket{\psi(t+dt)}=U(t+dt,t)\ket{\psi(t)}
\end{equation}
where the time evolution operator $U(t+dt,t)$ is given by the time-ordered integral
\begin{equation}
    U(t+dt,t)=\tau \exp\left[-i \int\limits_{t}^{t+dt}H(\tau)d\tau)\right],
\end{equation}
here $\tau$ is the time ordering operator. By using second order Trotter-Suzuki formula we can decompose the propagator $U(t+dt,t)$ as \cite{Wiebe2011,Poulin2015, qgs_sup_NielsenCUP2000, Daskin2011}
\begin{align}
    U(t+dt,t)&\approx e^{-i\frac{V}{2} dt} e^{-i E(t+\frac{dt}{2}) \frac{dt}{2}} e^{-i T dt} \nonumber\\
    &\quad\times e^{-i E(t+\frac{dt}{2}) \frac{dt}{2}} e^{-i\frac{V}{2} dt}.
\end{align}

Since the kinetic energy operator is diagonal in momentum basis, to make the simulation simpler, we perform quantum Fourier transform ($QFT$) . So the modified unitary evolution operator takes the following form
\begin{equation}
    U(t+dt,t)\approx \mathbb{V}_{dt/2} \mathbb{E}_{dt/2} (QFT) \mathbb{T}_{dt} (QFT)^{-1} \mathbb{E}_{dt/2} \mathbb{V}_{dt/2}
\end{equation}
\begin{figure}[H]
   \begin{center}
   \includegraphics[width=\linewidth]{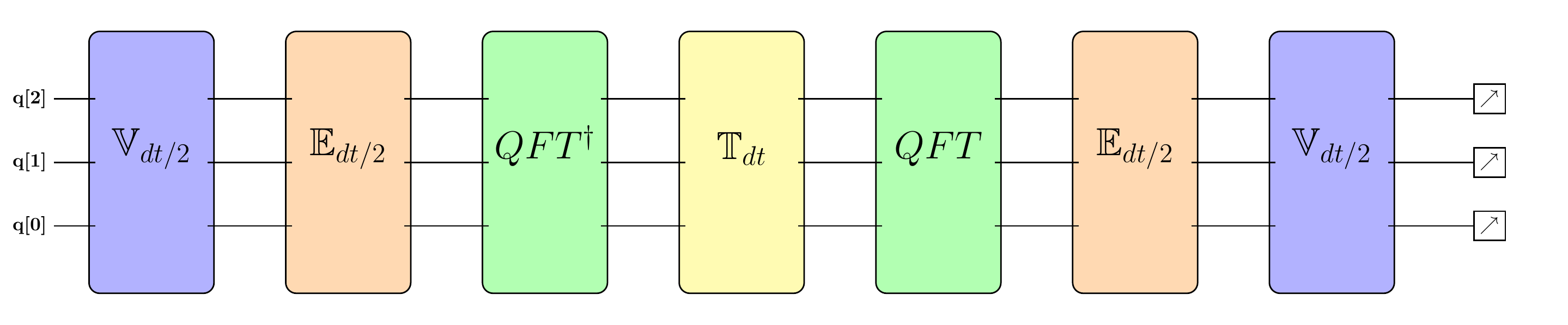}
   \end{center}
  \caption{\textbf{Quantum circuit for implementing laser driven isomerization reaction on a 3 qubit system}. Only one time step evolution of the reaction mechanism is illustrated here.}
  \label{circuit}
\end{figure}

where the operators
\begin{align}
    \mathbb{V}_{dt/2}=& e^{-iV\frac{dt}{2}}\nonumber\\
    \mathbb{T}_{dt}=& e^{-iTdt}\nonumber\\
    \mathbb{E}_{dt/2}=& e^{iE\frac{dt}{2}}\nonumber
\end{align}

In our experiments, we have considered a three qubit system to simulate the reaction dynamics. For that we discretize the space on $2^3=8$ lattice points with spacing $\Delta x$, and we encode the wave function on quantum registers as
\begin{align}
    \ket{\psi(x,t)}=&\displaystyle\sum_{l=0}^{2^n-1}\psi(x_l,t)\ket{l}\\
    =\psi(x_0,t)&\ket{000}+\psi(x_1,t)\ket{001}+...+\psi(x_{7},t)\ket{111}.\nonumber
\end{align}

Similarly, we divide the total time into 20 small time steps with interval $dt=3.75$. We show the reaction mechanism in three subsequent stages: (a) the initial time interval where the laser pulse is switched on, (b) the intermediate interval where the field is kept constant (c) the final stage where the field is turned off gradually. 
The behaviour of the electric field is given by  \cite{Lu2011}
\[\epsilon(t)=
\begin{cases}
    \epsilon_0 \sin^2 \left(\frac{\pi t}{2 \uptau_1}\right) &\quad 0\leqslant t \leqslant \uptau_1\\
    \epsilon_0 &\quad \uptau_1\ \textless\ t\ \textless\ \uptau_2\\
    \epsilon_0 \sin^2 \left[ \frac{\pi (t_f-t)}{2 (t_f-\uptau_2)} \right] &\quad \uptau_2\leqslant t \leqslant t_f
\end{cases}
\]
For $\epsilon_0=1\times 10^{-3}$, the electric field at different time steps takes the values
\begin{align*}
    \epsilon(t)=&[0.025,\ 0.206,\ 0.500,\ 0.794,\ 0.975,\ 1,\ ...1,\ 0.975,\\
    &0.794,\ 0.500,\ 0.206,\ 0.025]\times 10^{-3}
\end{align*}
Since we discretized the space using 8 grid points, the diagonal operators are given by
\begin{align*}
    V=&\ \textrm{diag}\ [293.78,\ -0.10,\ 1.85,\ 5.41,\ 5.46,\ 2.02,\\
    &0.18,\ 305.44]\times10^{-3}\\
    T=&\ \textrm{diag}\ [0.00,\ 0.91,\ 3.63,\ 8.16,\ 14.51,\ 8.16,\ 3.63,\\ &-0.91]\times10^{-3}\\
    x=&\ \textrm{diag}\ [-1.51,\ -1.08,\ -0.65,\ -0.22,\ 0.22,\ 0.65,\\
    &1.08,\ 1.51]
\end{align*}
In Fig. \ref{circuit}, we have shown the circuit implementation for one time step evolution of the chemical reaction. We make use of Walsh-series approximation \cite{Welch2014} for implementing diagonal unitary operators without ancillas. The detailed method is explained in the supplemental material.
Initially we prepare our state in $\ket{\psi_r}=\ket{010}$, which corresponds to the ground state of the reactant. During the time evolution we apply the electric field in such a way that the final state should approach the product state $\ket{\psi_p}$. After each time step we make the measurement to keep track of the reaction process.\\

\begin{figure}[H]
   \begin{center}
   \includegraphics[width=\linewidth]{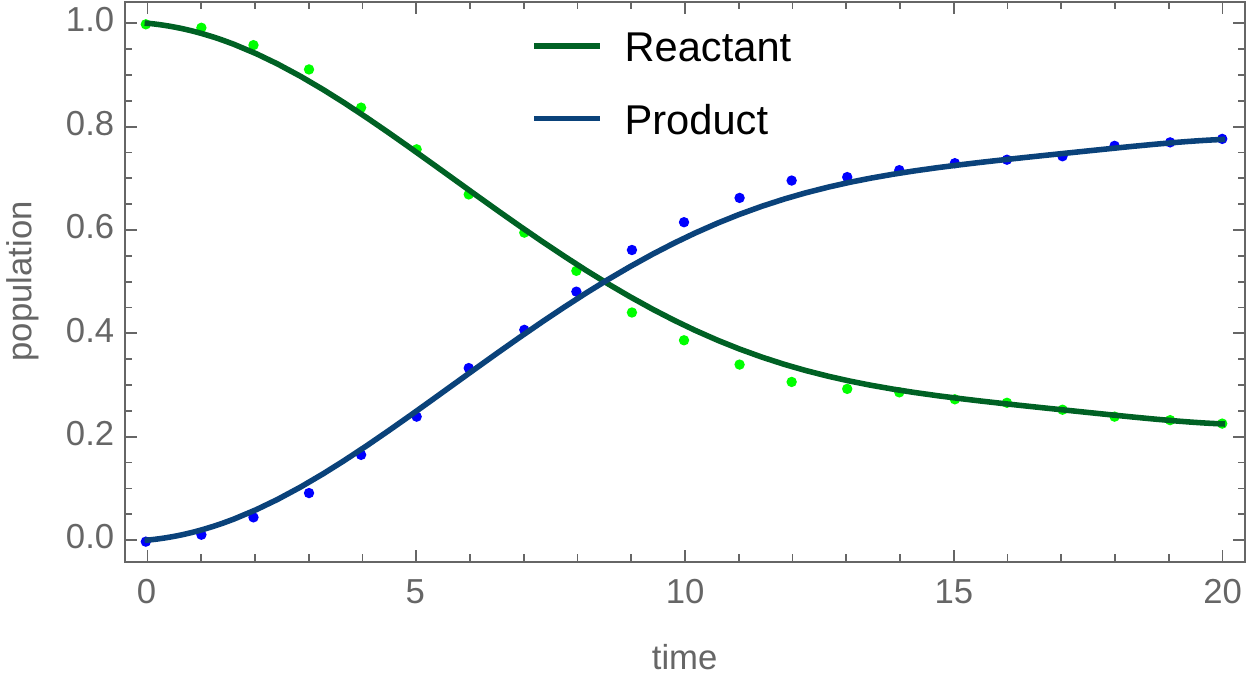}
   \caption{The population of the reactant state (green) and the product state (blue) in each time steps during the reaction process is depicted. As the time evolves, the reactant is gradually converted into product.}
   \label{population}
   \end{center}
\end{figure}

\begin{figure}
   \begin{center}
   \includegraphics[width=\linewidth]{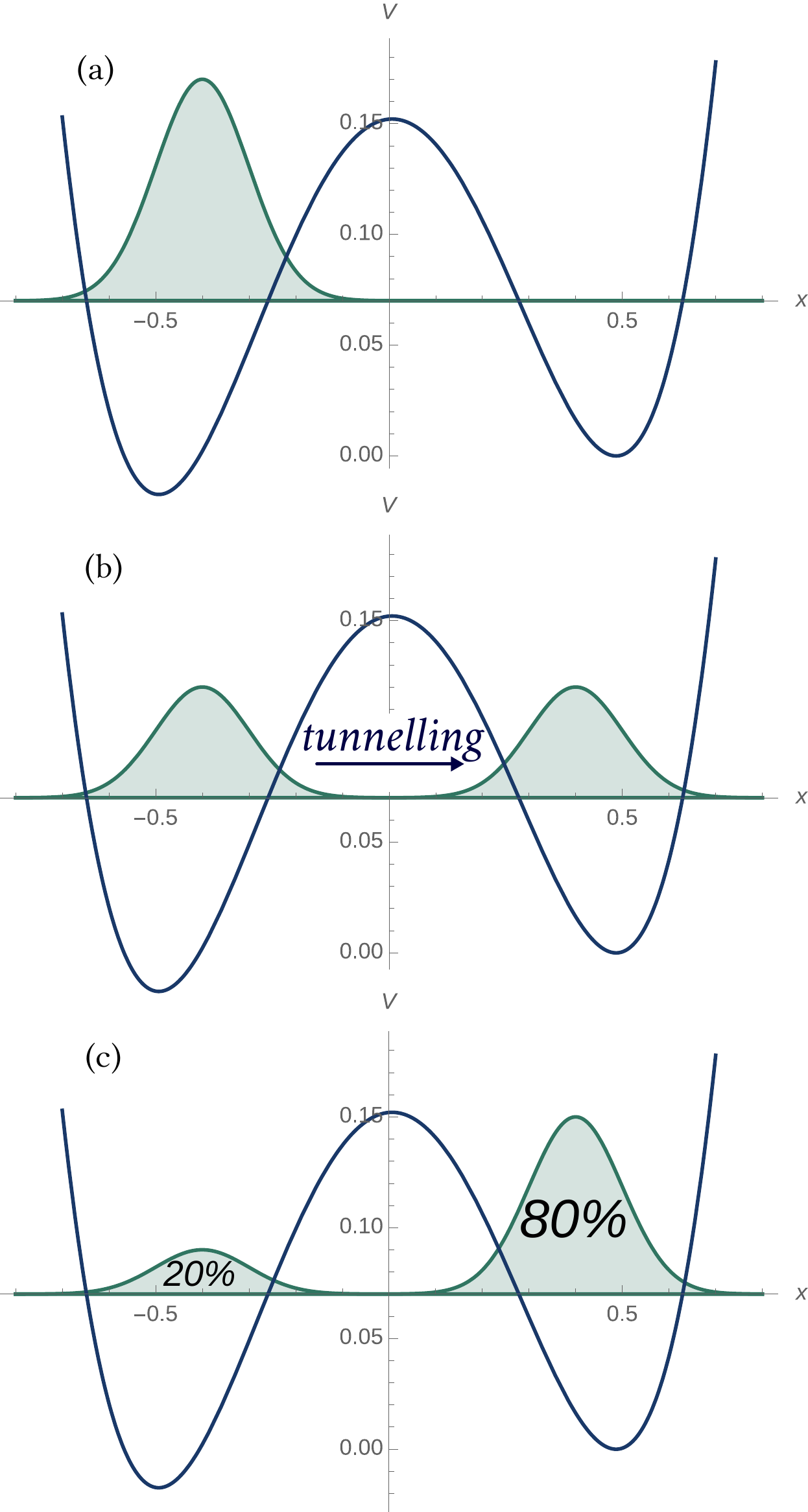}
   \caption{\textbf{Reaction mechanism in three different time intervals:} (a) Initially we apply the laser pulse such that the reactant starts converting into super position of near-degenerate delocalised states. (b) In the second time domain, the applied laser field is kept constant so that the wave packet tunnels through the barrier from reactant state to product state. (c) In the final stage we turn off the laser field so that we will get the product state.  }
   \label{tunnel}
   \end{center}
   \end{figure}

\onecolumngrid
\ 
\begin{figure}[H]
   \begin{center}
   \includegraphics[width=\textwidth]{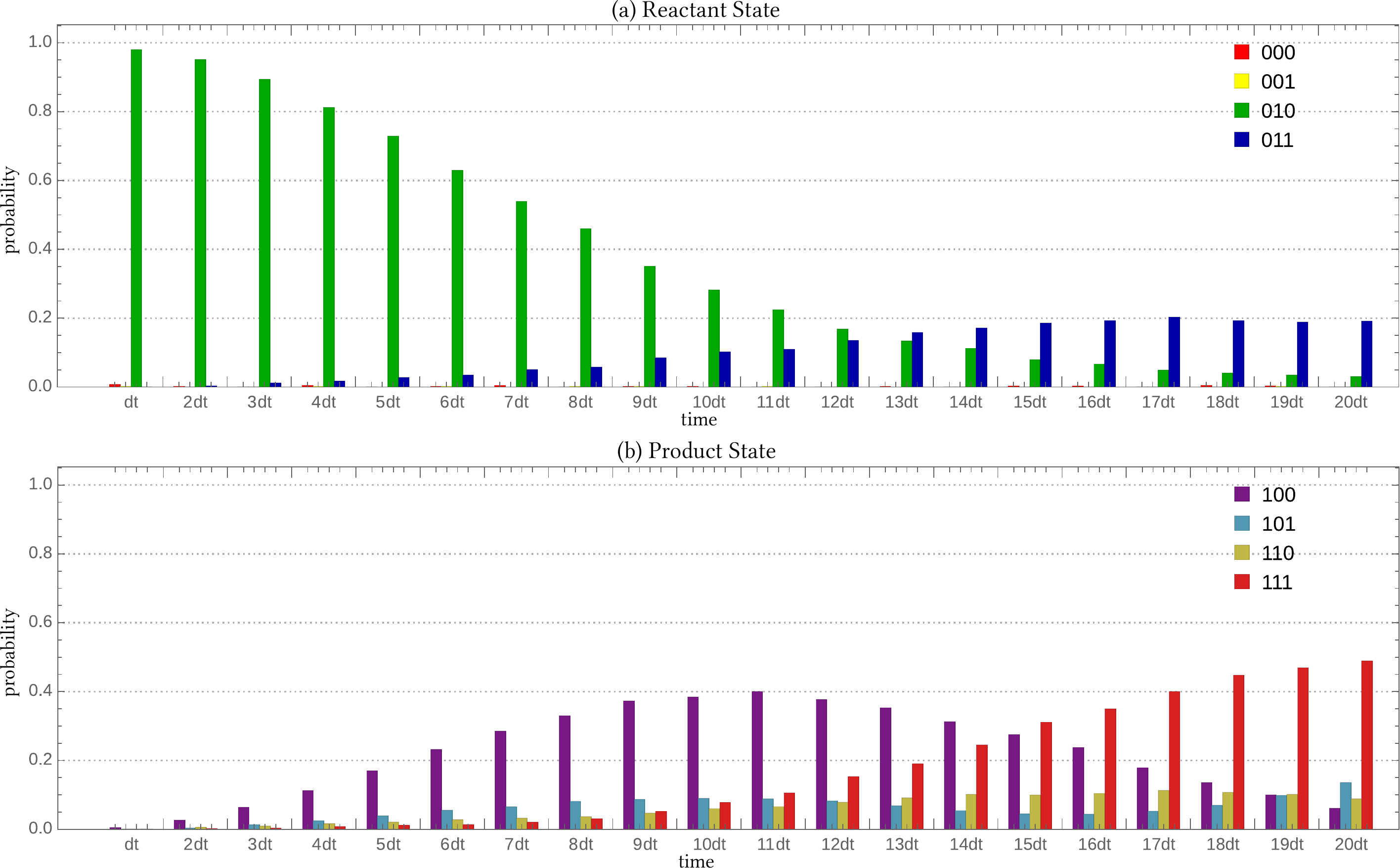}
    \caption{\textbf{The measured probabilities of the reactant state and the product state :} (a) The decrease in population of the reactant state with increasing time step is depicted. (b) As the time evolves the reactant is converted into product. At about $t=10 dt$, there is significant probability of finding both the reactant and product in $\ket{011}$ and $\ket{100}$ states respectively, which are between the potential barrier. This confirms that this reaction is essentially a tunneling phenomena.}
    \label{prob}
    \end{center}
    \end{figure}
\twocolumngrid
\emph{\textbf{Results:}} In Fig. \ref{tunnel} we have illustrated the reaction process in three different time intervals. Initially the reactant is localized in left potential well, by introducing the laser pulse the reactant starts converting into superposition of near-degenerate delocalized state. In the intermediate stage constant electric filed is applied, the wave packet representing the reactant starts to tunnels through the barrier. At the final stage we turn off the laser pulse, which results in the stabilization of wave function as product state in the right potential well.
\\The probabilities of reactant and product state during twenty time step evolution is shown in Fig. \ref{prob}. The continuous transformation of reactant to product state is clearly observed from the Fig. \ref{population}.  At the end of the reaction the population of the product state reaches 78\%.\\

The simulation has been performed in IBM's ideal quantum simulator (\emph{ibmq\_qasm\_simulator}) using 3 qubits. Since the number of gate operations are very large, using the quantum processor was impractical because of accumulation of gate errors.

\emph{\textbf{Conclusion:}} 
To summarize, we have demonstrated quantum simulation of laser-driven chemical isomerization reaction in its entirety in 20 discrete time steps. In each time step we have discretized the wave function in position basis to keep track of the product yield and the progress of the reaction. The formation of product from reactant can be better understood as a tunneling phenomena through a potential barrier. The final state and the yield of product is in strong agreement with theoretical predictions.

Efficient simulation of polyatomic reactions is one of the biggest objective of quantum chemistry and quantum simulation. The method we have used to implement arbitrary diagonal operators in a quantum circuit do not require any ancillary qubits. Further, depending on the degree of discretization, the circuit can be easily scaled to any number of qubits.

Accurate simulation of reaction dynamics over a long stretch of time requires quantum processors that have high coherence time and low gate and readout error. All of the currently available open quantum computing platforms (IBM, Riggeti, etc.) have high gate errors that prohibit precise simulation of multi-molecular reactions for a long period of time. With proper error-correction, this simulation can be run on a quantum processor in future.

\par
{\em \textbf{Author contributions:}} Theoretical analysis, design of quantum circuit and simulation were performed by N.N.H. and K.H. Collection and analysis of data was done by N.N.H. and K.H. The project was supervised by B.K.B. P.K.P. has thoroughly reviewed the manuscript. K.H., N.N.H. and B.K.B. have completed the project under the guidance of P.K.P.

{\em \textbf{Acknowledgements:}} We thank Indian Institute of Science Education and Research Kolkata for providing hospitality during which this work was completed. We acknowledge the support of IBM Quantum Experience for providing access to the quantum processors. The discussions and opinions developed in this paper are only those of the authors and do not reflect the opinions of IBM or any of it's employees.

\onecolumngrid
\newpage
\section*{SUPPLEMENTAL MATERIAL: Digital Quantum Simulation of Laser-Pulse Induced Tunneling Mechanism in Chemical Isomerisation Reaction}\label{sec_sup}
\twocolumngrid

To implement an arbitrary diagonal unitary operator using quantum circuit we need to decompose it into a sequence of at most two-level unitary gates. This is an ardous task for most operators that we encounter in nature \cite{Daskin2011}.  In this paper, we apply a method given by Welch et al. \cite{Welch2014} using Walsh function and Walsh operators without requiring any ancillary qubits. We demonstrate this method in $n=3$ qubits for one of the operators $V$.

The discrete Walsh-Hadamard function (or Walsh function) is sampled by discretizing the interval $[0,1)$ into $N=2^n$ points, giving us $x_j=j/N$ where $j=0,1,...,N-1$. Each element of the Walsh function can be obtained from
\begin{equation}
    w_{ij}=(-1)^{\sum_{k=1}^{n} i_k x_k},
\end{equation}
which for $n=3$, gives us a $8\times8$ matrix
\begin{equation}
    W_8=
    \begin{bmatrix}
    \ \ 1&\ \ 1&\ \ 1&\ \ 1&\ \ 1&\ \ 1&\ \ 1&\ \ 1\\
    \ \ 1&\ \ 1&\ \ 1&\ \ 1&-1&-1&-1&-1\\
    \ \ 1&\ \ 1&-1&-1&-1&-1&\ \ 1&\ \ 1\\
    \ \ 1&\ \ 1&-1&-1&-1&-1&\ \ 1&\ \ 1\\
    \ \ 1&\ \ 1&-1&-1&\ \ 1&\ \ 1&-1&-1\\
    \ \ 1&-1&-1&\ \ 1&-1&\ \ 1&\ \ 1&-1\\
    \ \ 1&-1&\ \ 1&-1&-1&\ \ 1&-1&\ \ 1\\
    \ \ 1&-1&\ \ 1&-1&\ \ 1&-1&\ \ 1&-1\\
    \end{bmatrix}
\end{equation}

For demonstration, we implement the operator as follows
\begin{equation}
\mathbb{V}_{dt/2}=e^{i\hat{f}_j}=e^{-i V dt/2}
\end{equation}

The discretized eigenfunction for this operator is
\begin{align}
    f_j=-V\times\frac{dt}{2}=&-[9.113,\ -0.003,\ 0.057,\ 0.168,\nonumber\\
    &0.169,\ 0.063,\ 0.006,\ 9.475]
\end{align}
The discrete Walsh-Fourier transform of $f_j$ is
\begin{equation}\label{WF}
    a_i=\frac{1}{N}\displaystyle\sum_{j=0}^{N-1} f_j w_{ij}
\end{equation}
The Walsh operator $\hat{w_i}$ corresponding to it is found from the reversed binary string of $i$. For instance, to find $\hat{w_4}$ on $n=3$ qubits, we bit-reverse $i=100$ (which is $001$) and replace the zeros by identity and ones by Pauli Z gates. This gives us $\hat{w_4}=\hat{I}\otimes\hat{I}\otimes\hat{Z}$.

Using these pairs of Walsh-Fourier functions and Walsh operators, any n qubit diagonal operator can be expressed as a sequence of $N=2^n$ exponential unitary operators given by
\begin{equation}
    \hat{U}=\displaystyle\prod_{i=0}^{N-1} \hat{U_i}=\displaystyle\prod_{i=0}^{N-1} e^{i a_i \hat{w_i}}
\end{equation}

Here the circuit for each operator $\hat{U_j}$ is a rotation gate about $Z$ axis given by $R_i \equiv R_Z(\theta_i)=e^{-iZ \theta_i /2}$ where $\theta_i=-2a_i$. To reduce number of redundant gates and gate error, we rearrange these operators in \emph{sequency order} \cite{Geadah1997}. Using Eq. \eqref{WF} we calculate the Walsh-Fourier coefficients and the corresponding rotations

\begin{center}
\begin{tabular}{c|rcc|r}
\cline{1-2} \cline{4-5} 
\multicolumn{2}{c}{$a_{i}$} & \hspace{1cm} & \multicolumn{2}{c}{$\theta_{i}=-2 a_i$}\tabularnewline
\cline{1-2} \cline{4-5} 
$a_{0}$ & -19.041 & \hspace{1cm} & $\theta_{0}$ & 38.083\tabularnewline
$a_{1}$ & 0.377 & \hspace{1cm} & $\theta_{1}$ & -0.754\tabularnewline
$a_{2}$ & -18.127 & \hspace{1cm} & $\theta_{2}$ & 36.254\tabularnewline
$a_{3}$ & 0.363 & \hspace{1cm} & $\theta_{3}$ & -0.727\tabularnewline
$a_{4}$ & -18.796 & \hspace{1cm} & $\theta_{4}$ & 37.593\tabularnewline
$a_{5}$ & 0.349 & \hspace{1cm} & $\theta_{5}$ & -0.698\tabularnewline
$a_{6}$ & -18.362 & \hspace{1cm} & $\theta_{6}$ & 36.725\tabularnewline
$a_{7}$ & 0.357 & \hspace{1cm} & $\theta_{7}$ & -0.713\tabularnewline
\cline{1-2} \cline{4-5} 
\end{tabular}
\end{center}

\begin{figure}[H]
   \begin{center}
   \includegraphics[width=\textwidth]{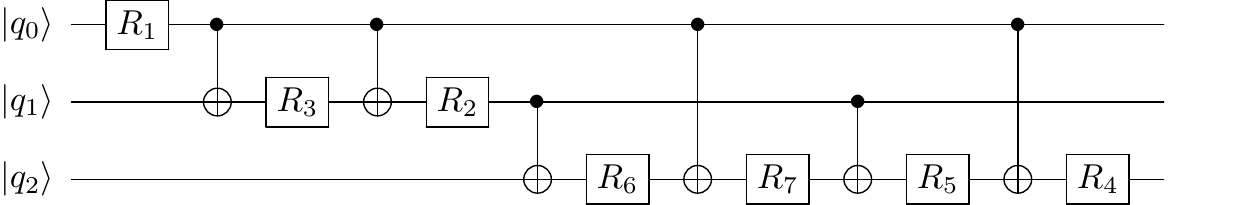}
    \caption{\textbf{Optimal circuit for implementing a diagonal unitary operator on a 3 qubit system:} the Walsh operators are sequency ordered which cancels CNOT gates between adjacent gates. The rotation gates about Z axis are $R_i\equiv R_Z(-2 a_i)$.}
    \label{walsh}
    \end{center}
    \end{figure}

Fig. \ref{walsh} shows the optimal quantum circuit that uses Walsh operators to implement any diagonal unitary operator for three qubits.

\onecolumngrid
\ 

\end{document}